\documentclass[conference]{IEEEtran}
\IEEEoverridecommandlockouts

\usepackage{amsmath,amssymb,amsfonts}
\usepackage{algorithmic}
\usepackage{graphicx}
\usepackage{textcomp}
\usepackage{xcolor}
\usepackage{hyperref}
\usepackage{glossaries}
\glsdisablehyper    

\usepackage{amsmath, amssymb, bm} 
\usepackage{amsthm}
\usepackage{dsfont}

\usepackage{tikz}
\usetikzlibrary{calc}
\usetikzlibrary{angles, quotes}
\usetikzlibrary{decorations.pathreplacing,decorations.markings,shapes.geometric}

\usepackage{comment}

\usepackage{siunitx}

\usepackage{stfloats}
\usepackage{float}

\usepackage{subcaption}  

\usepackage{cancel}
\usepackage[normalem]{ulem}

\usepackage{makecell}

\hypersetup{
    colorlinks=true,
    linkcolor=black,
    citecolor=black,
    urlcolor=black
}

\definecolor{myBlack}{HTML}{565656}
\definecolor{myGreen}{HTML}{5bb900}
\definecolor{myBlue}{HTML}{005ae9}
\definecolor{myRed}{HTML}{d7001d}
\definecolor{myOrange}{HTML}{f2a202}
\definecolor{myPurple}{HTML}{6000e2}
\definecolor{myPink}{HTML}{F03080}

\newtheorem{assumption}{Assumption}

\newcommand{\UErange}{R_{\mathrm{s}}}
\newcommand{\UEaoa}{\theta_{\mathrm{s}}}
\newcommand{\UErangeTest}{\widetilde{R}_{\mathrm{s}}}
\newcommand{\UEaoaTest}{\widetilde{\theta}_{\mathrm{s}}}
\newcommand{\UErangeEst}{\widehat{R}_{\mathrm{s}}}
\newcommand{\UEaoaEst}{\widehat{\theta}_{\mathrm{s}}}
\newcommand{\UErangeRes}{\Delta_{R_{\mathrm{s}}}}
\newcommand{\UEsinaoaRes}{\Delta_{\sin(\theta_{\mathrm{s}})}}

\newcommand{\channelMat}{\boldsymbol{H}}

\newcommand{\channelcoeff}{\gamma}

\newcommand{\channelcoeffTest}{\widetilde{\gamma}}
\newcommand{\channelcoeffEst}{\widehat{\gamma}}
\newcommand{\staoa}{\boldsymbol{a}}
\newcommand{\strange}{\boldsymbol{b}}
\newcommand{\stcomb}{\boldsymbol{v}}
\newcommand{\noiseVar}{\sigma_{\mathrm{n}}^{2}}
\newcommand{\channelMatEst}{\boldsymbol{Y}_{\mathrm{P}}}
\newcommand{\channelVecEst}{\boldsymbol{y}_{\mathrm{P}}}

\newcommand{\snr}{\mathrm{SNR}}

\newcommand{\pilotObs}{\boldsymbol{\mathcal{Y}}_{\mathrm{P}}}
\newcommand{\dataObs}{\boldsymbol{\mathcal{Y}}_{\mathrm{D}}}

\newcommand{\pilotAWGN}{\boldsymbol{\mathcal{N}}_{\mathrm{P}}}
\newcommand{\dataAWGN}{\boldsymbol{\mathcal{N}}_{\mathrm{D}}}

\newcommand{\pilotMat}{\boldsymbol{S}_{\mathrm{P}}}
\newcommand{\dataMat}{\boldsymbol{S}_{\mathrm{D}}}

\newcommand{\dataTest}{\widetilde{\boldsymbol{S}}_{\mathrm{D}}}
\newcommand{\dataEst}{\widehat{\boldsymbol{S}}_{\mathrm{D}}}
\newcommand{\dataConst}{\mathcal{C}}
\newcommand{\dataConstVar}{\sigma_{\mathrm{s}}^{2}}

\newcommand{\numP}{P}
\newcommand{\numD}{D}
\newcommand{\numQ}{Q}
\newcommand{\BW}{B}
\newcommand{\Time}{T}
\newcommand{\numRX}{N}
\newcommand{\RXspacing}{\Delta_{\mathrm{d}}}
\newcommand{\Qspacing}{\Delta_{\mathrm{f}}}
\newcommand{\Tspacing}{\Delta_{\mathrm{t}}}
\newcommand{\carrierF}{f_{\mathrm{c}}}
\newcommand{\carrierWl}{\lambda_{\mathrm{c}}}
\newcommand{\carrierWn}{k_{\mathrm{c}}}

\newcommand{\pilotMethod}{\textsc{Pilot}}
\newcommand{\pilotdataMethod}{\textsc{Genie}}
\newcommand{\projMethod}{\textsc{JPUDL}}

\newcommand{\ddlmmseMethod}{\textsc{DD (LMMSE)}}

\DeclareMathOperator{\vecc}{vec}

\DeclareMathOperator*{\argmin}{argmin}
\DeclareMathOperator*{\argmax}{argmax}

\newcommand{\norm}[1]{\left\lVert #1 \right\rVert_{\mathrm{2}}}
\newcommand{\abs}[1]{\left\lvert #1 \right\rvert}

\newcommand{\Compl}[1]{\mathcal{O}\left(#1\right)}
\newcommand{\numRXgrid}{N_{\mathrm{g}}}
\newcommand{\numQgrid}{Q_{\mathrm{g}}}

\newcommand{\CN}{\mathcal{CN}}

\newacronym{ofdm}{OFDM}{Orthogonal Frequency-Division Multiplexing}
\newacronym{isac}{ISAC}{Integrating Sensing And Communications}
\newacronym{bs}{BS}{Base Station}
\newacronym{ula}{ULA}{Uniform Linear Array}
\newacronym{ue}{UE}{User Equipment}
\newacronym{aoa}{AoA}{Angle Of Arrival}
\newacronym{los}{LoS}{Line of Sight}
\newacronym{ff}{FF}{Far-Field}
\newacronym{nf}{NF}{Near-Field}
\newacronym{pr}{PR}{Passive Radar}
\newacronym{awgn}{AWGN}{Additive White Gaussian Noise}
\newacronym{ml}{ML}{Maximum Likelihood}
\newacronym{ls}{LS}{Least Squares}
\newacronym{psk}{PSK}{Phase-Shift Keying}
\newacronym{zf}{ZF}{Zero Forcing}
\newacronym{snr}{SNR}{Signal-to-Noise Ratio}
\newacronym{fft}{FFT}{Fast Fourier Transform}
\newacronym{dd}{DD}{Decision-Directed}
\newacronym{lmmse}{LMMSE}{Linear Minimum Mean Square Error}
\newacronym{rmse}{RMSE}{Root Mean Square Error}
\newacronym{ser}{SER}{Symbol Error Rate}
\newacronym{prs}{PRS}{Positioning Reference Signal}
\newacronym{elmmse}{ELMMSE}{Ergodic Linear Minimum Mean Square Error}
\newacronym{tx}{TX}{transmitting}
\newacronym{rx}{RX}{receiving}
\newacronym{srx}{SRX}{Sensing Receiver}
\newacronym{6g}{6G}{Sixth Generation}
\newacronym{5g-nr}{5G-NR}{Fifth Generation New Radio}

\usepackage{fancyhdr}
\def\BibTeX{{\rm B\kern-.05em{\sc i\kern-.025em b}\kern-.08em
    T\kern-.1667em\lower.7ex\hbox{E}\kern-.125emX}}
\begin{document}

\title{Joint Pilot and Unknown Data-based Localization \\ for OFDM Opportunistic Radar Systems\\
\thanks{Mathieu Reniers is a Research Fellow of the Fonds de la Recherche
Scientifique - FNRS.
}
}

\author{
    Mathieu Reniers, 
    Martin Willame, 
    Jérôme Louveaux, 
    Luc Vandendorpe, \\
    ICTEAM, UCLouvain - Louvain-La-Neuve, Belgium. \footnotesize{Emails: \{firstname.lastname\}@uclouvain.be}
}

\bstctlcite{MyBSTcontrol}

\maketitle
\thispagestyle{fancy}

\begin{abstract}
Integrating Sensing and Communications (ISAC) has emerged as a promising paradigm for Sixth Generation (6G) and Wi-Fi 7 networks, with the communication-centric approach being particularly attractive due to its compatibility with current standards. Typical communication signals comprise both deterministic known pilot signals and random unknown data payloads. Most existing approaches either rely solely on pilots for positioning, thereby ignoring the radar information present in the received data symbols that constitute the majority of each frame, or rely on data decisions, which bounds positioning performance to that of the communication system. To overcome these limitations, we propose a novel method that extracts positioning information from data payloads without decoding them. We consider an opportunistic scenario in which communication signals from a user are captured by a passive radar equipped with a uniform linear array of antennas. We show that, in this setting, the estimation can be efficiently implemented using Fast Fourier Transforms. Finally, we demonstrate superior localization performance compared to existing methods in the literature through numerical simulations.
\end{abstract}
\glsresetall

\begin{IEEEkeywords}
Integrated Sensing and Communication, Data-aided Localization, OFDM Passive Radar, WiFi Sensing, Fast Fourier Transform
\end{IEEEkeywords}

\vspace{-0.25cm}
\section{Introduction}
Recent years have witnessed growing interest in \gls{isac} systems as a paradigm for both future \gls{6g} networks and new Wi-Fi 7 systems.
Integrating both services improves spectrum efficiency \cite{liu_joint_2020} and enables hardware sharing \cite{feng_joint_2020}, thereby reducing device cost, size, and power consumption while potentially enhancing both communication and sensing performance \cite{zhang_overview_2021}. 
Due to its compatibility with existing wireless communication standards, the \textit{communication-centric} approach—where communication remains the primary function, and sensing or radar capabilities are performed using communication signals—has been investigated and appears most promising \cite{lu_sensing_2026}.
A communication signal comprises both pilots or reference signals alongside data payloads.
The pilots are \textbf{deterministic} and \textbf{known} by both \gls{tx} and \gls{rx} sides while the \textbf{random} data payload conveys information and is \textbf{unknown} at the \gls{rx}.

Localization based solely on pilots has been widely studied in the literature, with several recent works focusing on optimal pilot design for \gls{isac} \cite{wei_5g_2023,golzadeh_joint_2024}.
Wei et al. \cite{wei_5g_2023} explored radar performance using the \gls{5g-nr} \gls{prs}.
This work was extended by Golzadeh et al. \cite{golzadeh_joint_2024}, who introduced irregular resource patterns in \gls{prs} to further enhance sensing performance in high-mobility environments. 
However, these pilot-based localization techniques completely overlook the positioning information contained in the data payloads, which occupy the majority of each transmitted frame (typically \SI{75}{\percent}$-$\SI{97}{\percent} in \gls{5g-nr} \cite{3gpp_5g_2025}).

Consequently, several studies have focused on characterizing sensing performance under random signals rather than deterministic pilots \cite{liu_cp-ofdm_2025,xu_exploiting_2026}. 
These works consider monostatic radar configurations, i.e., colocated \gls{tx} and \gls{srx}, where perfect knowledge of the random data symbols at the \gls{srx} is assumed.
Liu et al. \cite{liu_cp-ofdm_2025} investigated optimal communication-centric ranging waveforms, demonstrating that \gls{ofdm} is the unique optimal waveform achieving the lowest ranging sidelobes.
Xu et al. \cite{xu_exploiting_2026} derived a semi-closed form of the \glsdesc{elmmse}, which quantifies the average sensing error when exploiting random data payload symbols, and proposed efficient precoding strategies based on these findings.
While these studies provide valuable insights into the impact of data randomness, perfect knowledge of the data at the \gls{srx} is not attainable in multistatic radar or opportunistic sensing configurations.

A natural way to exploit random unknown data symbols is to decode them, following a \gls{dd} philosophy. 
This approach has been extensively integrated into modern \gls{isac} systems, often through recursive processing schemes that create a beneficial cycle between \textit{communication-aided sensing} and \textit{sensing-aided communication} \cite{zhao_joint_2024,keskin_bridging_2025-1}.
Zhao et al. \cite{zhao_joint_2024} developed this framework for single-carrier bistatic configurations, while Keskin et al. \cite{keskin_bridging_2025-1} extended it to \gls{ofdm}.
While these methods can achieve very low localization errors under favorable \gls{snr} conditions, they suffer from high computational complexity due to data decoding and iterative processing, and their localization performance is inherently limited by the communication performance.

To address these limitations, we propose a novel method that leverages both pilots and data payloads for positioning without requiring data decoding, enabling operation in low \gls{snr} conditions.
We consider an uplink scenario in which the \gls{tx} is a \gls{ue} transmitting communication signals to a \gls{rx}.
These signals are simultaneously received by the \gls{srx}—an opportunistic \gls{pr} in this context—whose objective is to localize the \gls{ue} without performing any communication-related task.

\begin{figure}[H]
    \centering
        \resizebox{0.92\linewidth}{!}{%
\tikzset{antenna/.style={insert path={-- coordinate (ant#1) ++(0,0.25) -- +(135:0.25) + (0,0) -- +(45:0.25)}}}
\tikzset{station/.style={draw,shape=dart,shape border rotate=90, minimum width=10mm, minimum height=10mm,outer sep=0pt,inner sep=3pt}}

\newcommand{\MBS}[2][1]{%
\begin{scope}[shift={#2},scale=#1,transform shape]
    \node[thick,station] (base) {};
    \draw[thick, line join=bevel] (base.100) -- (base.80) -- (base.110) -- (base.70) -- (base.north west) -- (base.north east);
    \draw[thick, line join=bevel] (base.100) -- (base.70) (base.110) -- (base.north east);

    \def\antennaspacingBS{0.2}   
    \def\numBSAnt{8}               
        
    \pgfmathsetmacro{\arraywidth}{(\numBSAnt-1)*\antennaspacingBS}
    \pgfmathsetmacro{\xoffset}{-\arraywidth/2}
    
    \antennaarray{(\xoffset, 0.8)}{\numBSAnt}{\antennaspacingBS}{0.3}
    
    \draw[thick] (\xoffset-0.014, 0.8) -- (\xoffset + \arraywidth+0.014, 0.8);
    
\end{scope}%
}

\newcommand{\antenna}[2][1]{
    \draw[thick] #2 -- ++(0,0.5*#1);
    \draw[thick] #2 ++(0,0.5*#1) -- ++(-0.25*#1,0.25*#1);
    \draw[thick] #2 ++(0,0.5*#1) -- ++(0.25*#1,0.25*#1);
}
\def\antennaheight{0.75}    
\def\antennaspacing{0.75}   
\def\numRXAnt{5}               

\newcommand{\antennaarray}[5][0]{
    \begin{scope}[shift={#2}, rotate=#1]
        \foreach \i in {0,...,\numexpr#3-1\relax} {
            \antenna[#5]{(\i*#4,0)}
            \coordinate (ant\i) at (\i*#4,0);
        }
    \end{scope}
    \pgfmathsetmacro{\arraycenterx}{(#3-1)*#4/2}
    \pgfmathsetmacro{\arrayendx}{(#3-1)*#4}
    \coordinate (array-center) at (\arraycenterx,0);
    \coordinate (array-end) at (#3*#4-#4,0);
}

\def\commCOLOR{myBlue}
\def\oppCOLOR{myRed}

\begin{tikzpicture}
    \def\ueX{4.5}  
    \def\ueY{4.0}   
    \coordinate (ue) at (\ueX,\ueY);
    \antennaarray{(0,0)}{\numRXAnt}{\antennaspacing}{1}
    \node[font=\small,  text width=2cm, align=center] at ($(array-center)+(0.0,\antennaheight+0.3)$) {\gls{srx}};
    \draw[thick, <->] (0,0) -- (\antennaspacing,0) node[pos=0.5,above]{$\RXspacing$};
    \draw[thick, <->] (0,-0.2) -- (\antennaspacing*\numRXAnt-\antennaspacing,-0.2) node[pos=1.0,right]{$(\numRX-1)\RXspacing$};

    \fill[black] (ue) circle (0.08);
    \node[above] at ($(ue)+(0,0.1)$) {TX};
    \draw[black] (ue) circle (0.7);
    \draw[black!60!white] (ue) circle (0.9);
    \draw[black!20!white] (ue) circle (1.1);
    \draw[black!10!white] (ue) circle (1.3);
    \draw[\oppCOLOR, thick, dashed] ($(array-end)+(0,\antennaheight)$) -- (ue) node[midway, left] {$\UErange$};  
    \draw[<-, \oppCOLOR, dashed, thick] ($(array-end)+(-\antennaspacing,\antennaheight)$) -- ($(ue)-(\antennaspacing,0)$)
    node[pos=0.5, above=0.1cm, sloped, fill=\oppCOLOR!10!white, fill opacity=0.8, text opacity=1, rounded corners,] {Opportunistic Link};  
    
    \draw[black] ($(array-end)+(0,\antennaheight)$) -- ++(0,1);
    \draw[\oppCOLOR, thick,->] ($(array-end)+(0,\antennaheight+0.5)$) 
        arc[start angle=90, end angle={atan2(\ueY-\antennaheight, \ueX-\arrayendx)}, radius=0.5]
        node[pos=0.75, above=0.05cm] {$\UEaoa$};
    


    \draw[thick, black!50, ->] ($(array-end)+(0,\antennaheight)$)--($(array-end)+(\antennaspacing*0.5,\antennaheight)$) node[pos=1.0,below] {$x$};
    \draw[thick,black!50, ->] ($(array-end)+(0,\antennaheight)$)--($(array-end)+(0,\antennaheight*1.5)$) node[pos=1.0,left] {$y$};

    \coordinate (bs) at (0.0,4.0);
    \MBS[1.0]{(bs)}
    \node[below=0.5cm] at (bs) {\gls{rx}};

    \draw[\commCOLOR, thick, <-,] ($(bs)+(\antennaspacing,\antennaspacing)$) -- (ue) 
    node[pos=0.45, above=0.1cm, sloped, fill=\commCOLOR!10!white, fill opacity=0.8, text opacity=1, rounded corners,] {Communication Link};

\end{tikzpicture}
    }
    \caption{Illustration of the considered scenario. The uplink signal transmitted by the \gls{ue} (i.e., the \gls{tx}) is received by the intended communication \gls{rx}, and is simultaneously captured by a \gls{pr} (i.e., the \gls{srx}).}
    \label{fig:scenario}
\end{figure}

Our main contributions can be summarized as follows:
\begin{itemize}
    \item We develop a novel localization method that extracts position information from both pilots and data symbols without requiring data detection, making it agnostic to communication performance. 
    The method is derived for an uplink transmission and its effectiveness is validated in an opportunistic scenario where no communication performance guarantees exist (i.e., at low \gls{snr}).
    \item We show that when a \gls{ula} is used at the \gls{srx}, the method admits an efficient implementation based on \glspl{fft}.
    \item Through Monte Carlo simulations, we demonstrate superior localization performance over methods from the literature. 
    Besides, we provide a comprehensive analysis of how system parameters affect performance.
    \item We analyze the computational complexity of the proposed method against the considered baselines, highlighting the impact of system parameters on computational requirements.
\end{itemize}

Scalars, vectors, matrices and tensors are respectively defined as $a$, $\boldsymbol{a}$, $\boldsymbol{A}$ and $\boldsymbol{\mathcal{A}}$.
Vectors and matrices that are function of parameters are written as $\boldsymbol{a}(\cdot)$ and $\boldsymbol{A}(\cdot)$.
The $i$-th element of $\boldsymbol{a}$, the $(i,j)$-th element of $\boldsymbol{A}$ and the $(i,j,k)$-th element of $\boldsymbol{\mathcal{A}}$ are indicated by $\boldsymbol{a}[i]$, $\boldsymbol{A}[i,j]$ and $\boldsymbol{\mathcal{A}}[i,j,k]$. 
The symbol “:” denotes the selection of all elements along a given dimension; e.g., $\boldsymbol{\mathcal{A}}[i,j,:]$ denotes the vector of elements of $\boldsymbol{\mathcal{A}}$ along the third dimension for indices $(i,j)$.
The complex conjugate, the Hermitian Transpose and the Frobenius norm are expressed as $\boldsymbol{A}^*$, $\boldsymbol{A}^H$ and $\norm{\boldsymbol{A}}$.
The real and complex sets are denoted by $\mathds{R}$ and $\mathds{C}$. 
The Kronecker product is represented by~$\otimes$.

\section{System Model}

\subsection{Scenario}

In this paper, we investigate the problem of localizing a single-antenna \gls{ue} that emits an \gls{ofdm} uplink signal, which is captured by a \gls{pr} equipped with a \gls{ula} of antennas.
An illustration of the scenario is provided in \autoref{fig:scenario}. 
The \gls{pr} comprises $\numRX$ antennas spaced by $\RXspacing$.
The \gls{ue} transmits $\numP$ pilot and $\numD$ data \gls{ofdm} symbols across $\numQ$ subcarriers, with a subcarrier spacing of $\Qspacing$.
The frequency associated with the first subcarrier is denoted by $\carrierF$, with corresponding wavelength $\carrierWl$ and wavenumber $\carrierWn \triangleq 2\pi \carrierF / c = 2 \pi / \carrierWl$, where $c$ is the speed of light.
The pilot symbols are represented by $\pilotMat \in \mathds{C}^{\numQ \times \numP}$ and may consist of arbitrary complex sequences.
The data symbols are represented by $\dataMat \in \dataConst^{\numQ \times \numD} \subset \mathds{C}^{\numQ \times \numD}$, where each element belongs to a constellation from the set $\dataConst$. 
In general, the chosen constellation may differ for each time instant $d$ and subcarrier $q$.
This constellation assignment is determined by the \gls{tx} and is typically specified in the preamble containing transmission metadata.
However, since the proposed method relaxes the constellation constraint, we keep this implicit and use the previous notation for simplicity.
Without loss of generality, the symbols are transmitted according to the resource grid shown in \autoref{fig:resource_grid}.
\begin{figure}[t]
    \centering
    \resizebox{0.83\linewidth}{!}{%
        \def\numPilots{4}
\def\numData{16}
\pgfmathsetmacro{\numTimeSlots}{\numPilots+\numData}   
\def\numSubcarriers{7}   
\def\cellSize{0.25}       

\begin{tikzpicture}

    \pgfmathsetmacro{\numPilotsMinusOne}{\numPilots-1}
    \pgfmathsetmacro{\numTimeSlotsMinusOne}{\numTimeSlots-1}
    \pgfmathsetmacro{\numSubcarriersMinusOne}{\numSubcarriers-1}
    
    \foreach \x in {0,...,\numPilotsMinusOne} {
        \foreach \y in {0,...,\numSubcarriersMinusOne} {
            \fill[black!10] (\x*\cellSize, \y*\cellSize) rectangle ({(\x+1)*\cellSize}, {(\y+1)*\cellSize});
        }
    }
    
    \foreach \x in {\numPilots,...,\numTimeSlotsMinusOne} {
        \foreach \y in {0,...,\numSubcarriersMinusOne} {
            \fill[black!20] (\x*\cellSize, \y*\cellSize) rectangle ({(\x+1)*\cellSize}, {(\y+1)*\cellSize});
        }
    }

    \foreach \x in {0,...,\numTimeSlots} {
        \draw[black!30] (\x*\cellSize, 0) -- (\x*\cellSize, \numSubcarriers*\cellSize);
    }
    \foreach \y in {0,...,\numSubcarriers} {
        \draw[black!30] (0, \y*\cellSize) -- (\numTimeSlots*\cellSize, \y*\cellSize);
    }

    \draw[->, thick] (-0.75*\cellSize, 0) -- (\numTimeSlots*\cellSize + \cellSize, 0) node[pos=0.4, below=0.5cm] {Time};
    \draw[->, thick] (0, -0.75*\cellSize) -- (0, \numSubcarriers*\cellSize + \cellSize) node[pos=0.5, left=1.5cm, rotate=90, anchor=center] {Frequency};

    \draw[<->, thick] (0, -\cellSize) -- (\numPilots*\cellSize, -\cellSize) node[midway, below] {$\numP\Tspacing$};
    \draw[<->, thick] (\numPilots*\cellSize, -\cellSize) -- (\numTimeSlots*\cellSize, -\cellSize) node[midway, below] {$\numD\Tspacing$};
    \draw[<->, thick] (-\cellSize,0) -- (-\cellSize, \numSubcarriers*\cellSize) node[midway, left] {$\numQ\Qspacing$}
    node[pos=0, left=0.1cm] {$\carrierF$};

    \pgfmathsetmacro{\pilotCenter}{\numPilots*\cellSize/2}
    \pgfmathsetmacro{\dataCenter}{(\numPilots + \numTimeSlots)*\cellSize/2}
    \pgfmathsetmacro{\verticalCenter}{\numSubcarriers*\cellSize/2}
    
    \node[] at (\pilotCenter, \verticalCenter) {$\pilotMat$};
    \node[] at (\dataCenter, \verticalCenter) {$\dataMat$};

    \draw[<->, thick]
    (\numTimeSlots*\cellSize - 2*\cellSize, \numSubcarriers*\cellSize + 0.75*\cellSize)
    --
    (\numTimeSlots*\cellSize - 1*\cellSize, \numSubcarriers*\cellSize + 0.75*\cellSize)
    node[midway, above] {$\Tspacing$};

    \draw[<->, thick]
    (\numTimeSlots*\cellSize + 0.75*\cellSize, \numSubcarriers*\cellSize - 2*\cellSize)
    --
    (\numTimeSlots*\cellSize + 0.75*\cellSize, \numSubcarriers*\cellSize - 1*\cellSize)
    node[midway, right] {$\Qspacing$};

\end{tikzpicture}
    }
    \caption{Illustration of the considered resource grid.}
    \label{fig:resource_grid}
\end{figure}

\vspace{0.1cm}
We make the following assumption regarding the scene:
\begin{assumption}\label{ass:stillness}
    The \gls{ue} and the background are assumed to be stationary during the frame duration $\Time \triangleq (\numP+\numD)\Tspacing$, with $\Tspacing$ being the \gls{ofdm} symbol duration. Hence, no Doppler effect is considered.
\end{assumption}
Finally, localization is performed by estimating the range $\UErange$ and \gls{aoa} $\UEaoa$ associated with the \gls{ue}.
\subsection{Channel Model}
To derive the channel model, we make the following assumptions:
\begin{assumption}\label{ass:los}
    Only \gls{los} propagation is considered.
\end{assumption}
\begin{assumption}\label{ass:synchro}
    Time synchronization is already achieved and local oscillators are matched (i.e., no carrier frequency offset) between the \gls{ue} and the \gls{pr}. 
\end{assumption}
\begin{assumption}\label{ass:FF}
    The \gls{ue} is assumed to be in the \gls{ff} regime w.r.t. the \gls{pr}, i.e., $\UErange > R_{\mathrm{F}}$, where $R_{\mathrm{F}} = 2(\numRX\RXspacing)^2/\carrierWl$ is the Fraunhofer distance.
\end{assumption}
\begin{assumption}\label{ass:bw}
    The bandwidth of the transmitted signal is much smaller than the carrier frequency, i.e., $\BW \triangleq \numQ \Qspacing \ll \carrierF$.
\end{assumption}

Following these assumptions, the channel matrix can be written as (see, for instance, \cite{yildirim_enabling_2024})
\begin{equation}
    \channelMat(\UErange, \UEaoa) = \staoa(\UEaoa) \cdot \channelcoeff \cdot \strange^T(\UErange) \quad \in \mathds{C}^{N \times Q} , 
\end{equation}
where $\channelcoeff \in \mathds{C}$ is the random channel coefficient, including the propagation loss and absolute phase $e^{-j\carrierWn \UErange}$, while 
$\staoa(\UEaoa) \in \mathds{C}^{\numRX \times 1}$ and $\strange(\UErange) \in \mathds{C}^{\numQ\times 1}$ are respectively the steering vector associated with the \gls{aoa} (space domain) and the range (frequency domain). These are defined as
\begingroup \small
\begin{align}
    \staoa(\UEaoa) & = \begin{bmatrix}
        1 &  e^{-j\carrierWn\RXspacing\sin(\UEaoa)} & \cdots & e^{-j\carrierWn\RXspacing\sin(\UEaoa) (N-1)}
    \end{bmatrix}^{T} , \\
    \strange(\UErange) & = \begin{bmatrix}
        1 & e^{-j\carrierWn \frac{\Qspacing}{\carrierF} \UErange} & \cdots & e^{-j\carrierWn \frac{\Qspacing}{\carrierF} \UErange (\numQ-1)}
    \end{bmatrix}^{T} .
\end{align}
\endgroup
Note that this model implicitly assumes that $\Time < T_{\mathrm{c}}$ and $\BW < B_{\mathrm{c}}$ where $T_{\mathrm{c}}$ and $B_{\mathrm{c}}$ are the coherence time and bandwidth, respectively. 
These conditions are directly satisfied by Assumptions \ref{ass:stillness} and \ref{ass:los}.
The linear phase shifts arise from the \gls{ula} configuration and \autoref{ass:FF}, while \autoref{ass:bw} allows neglecting the cross-term $e^{-j\carrierWn n \RXspacing \sin(\UEaoa) q \Qspacing / \carrierF}$ that appears when expanding the channel expression.

At the \gls{pr} side, the observations associated with pilot and data components are respectively denoted as $\pilotObs \in \mathds{C}^{\numRX \times \numQ \times \numP}$ and $\dataObs \in \mathds{C}^{\numRX \times\numQ \times \numD}$ and are expressed as
\begin{align}
    \pilotObs[n,q,p] &= \channelMat(\UErange, \UEaoa)[n,q] \pilotMat[q,p] + \pilotAWGN[n,q,p] , \\
    \dataObs[n,q,d] &= \channelMat(\UErange, \UEaoa)[n,q] \dataMat[q,d] + \dataAWGN[n,q,d] ,
\end{align}
where $\pilotAWGN \in \mathds{C}^{\numRX \times \numQ \times \numP }$ and $\dataAWGN \in \mathds{C}^{\numRX \times  \numQ \times \numD}$ are the \gls{awgn} terms that are independent across all dimensions, with each element following a complex normal distribution $\CN(0,\noiseVar)$.

\section{Proposed Method}
The \gls{pr} knows the pilot sequences $\pilotMat$, as defined in communication standards, but the data symbols $\dataMat$ are unknown.
An intuitive approach to extract position information from the data observations is to demodulate the data symbols, following a \gls{dd} philosophy.
However, in an opportunistic scenario, information required for data demodulation (e.g., constellation assignment) is not always available.
Moreover, in typical systems, communication is optimized to ensure good performance between the \gls{tx} and the intended \gls{rx} (i.e., high \gls{snr} and low \gls{ser} on data payloads), while the \gls{pr} receives signals through an opportunistic link, where such favorable conditions cannot be guaranteed.
This motivates the need for efficient localization methods that do not rely on good communication performance.
This section derives the proposed method, which leverages both pilot and data observations without requiring data decoding, thereby avoiding its computational overhead while enabling operation under low \gls{snr} conditions.

\subsection{Maximum Likelihood Estimator}
In this section, we formulate the localization problem as a joint \gls{ml} and derive the corresponding estimator. 
The \gls{ml} estimation problem is expressed as
\begingroup \small
\begin{align}
    \UErangeEst, \UEaoaEst 
        &= \argmax_{\UErangeTest, \UEaoaTest}
            \max_{\dataTest \in \dataConst^{\numQ \times \numD}}
            \max_{\channelcoeffTest \in \mathds{C}}
            \;\mathcal{L}(\pilotObs,\dataObs ; 
              \UErangeTest, \UEaoaTest, \dataTest, \channelcoeffTest),
    \\
     &= \argmax_{\UErangeTest, \UEaoaTest}
            \max_{\dataTest \in  \dataConst^{\numQ \times \numD}}
            \max_{\channelcoeffTest \in \mathds{C}}
            \; \Big[
                \mathcal{L}(\pilotObs; 
                    \UErangeTest, \UEaoaTest, \channelcoeffTest)
                \nonumber \\
        &\hspace{3.5cm}\cdot\;
            \mathcal{L}(\dataObs;
                \UErangeTest, \UEaoaTest, \dataTest, \channelcoeffTest)
            \Big], \label{eq:ML_sep}
\end{align}
\endgroup
where $\mathcal{L}(\pilotObs, \dataObs ; \UErangeTest, \UEaoaTest, \dataTest, \channelcoeffTest)$ is the likelihood or probability of observing $(\pilotObs, \dataObs)$ given the \textit{parameters of interest} $(\UErangeTest, \UEaoaTest)$ and the \textit{nuisance parameters} $(\dataTest, \channelcoeffTest)$.
Equation \eqref{eq:ML_sep} is obtained by exploiting the independence between pilot and data observations to decompose the likelihood.

\subsubsection{Pilot component}
It can be shown, by expanding $\mathcal{L}(\pilotObs; \UErangeTest, \UEaoaTest, \channelcoeffTest)$, that it is strictly equivalent to work with $\mathcal{L}(\channelMatEst; \UErangeTest, \UEaoaTest, \channelcoeffTest)$ based on the transformed observation model $\channelMatEst \in \mathds{C}^{\numRX \times \numQ}$ defined as
\begingroup \small
\begin{equation}  \label{eq:channelEst}
        \channelMatEst[n,q] \triangleq \frac{\pilotMat[q,:]^{H} \pilotObs[n,q,:]}{\norm{\pilotMat[q,:]}^2}
     = \channelMat[n,q] + \frac{\pilotMat[q,:]^{H} \pilotAWGN[n,q,:]}{\norm{\pilotMat[q,:]}^2}.
\end{equation}
\endgroup
Intuitively, this simply involves equalizing the known pilots and using this \gls{zf} channel estimate instead of the raw observations.
This change of observation model is introduced as it allows simpler mathematical derivations and a more interpretable formulation, but its solution is exactly the same as that of the original \gls{ml}.
For unit-norm pilot symbols, $\norm{\pilotMat[q,:]}^2 = \numP$ and we have $\channelMatEst[n,q] \sim \CN(\channelMat[n,q], \noiseVar/\numP)$.
For simplicity in further derivations, we introduce $\channelVecEst = \vecc\{\channelMatEst\} \in \mathds{C}^{\numRX\numQ \times 1}$ and $\stcomb(\UErangeTest, \UEaoaTest) = \staoa(\UEaoaTest) \otimes \strange(\UErangeTest) \in \mathds{C}^{\numRX\numQ \times 1}$.

\subsubsection{Data component}
To derive our method, we \textbf{relax the constellation constraint}, i.e., we neglect the discrete nature of the data symbols, and maximize \eqref{eq:ML_sep} in $\dataTest \in \mathds{C}^{\numQ \times \numD}$ rather than $\dataTest \in \dataConst^{\numQ \times \numD}$.
This relaxation allows us to combine both nuisance parameters $(\dataTest, \channelcoeffTest)$ in the data component into a single one, $\dataTest' = \channelcoeffTest \dataTest \in \mathds{C}^{\numQ \times \numD}$, so that both terms in \eqref{eq:ML_sep} are affected by their own independent nuisance parameter.
Additionally, as each \textbf{random} and \textbf{unknown} transmitted symbol $\dataMat[q,d]$ produces only $\numRX$ coherent observations at the \gls{pr}, the data part provides information solely about the \gls{aoa}.
Specifically, no range information can be retrieved because there is no common structure across the frequency axis of $\dataMat$. 
Consequently, the observation model is rewritten as
\begin{equation}\label{eq:model_data_prime}
    \dataObs[n,q,d] = \staoa[n] \dataMat''[q,d] + \dataAWGN[n,q,d],
\end{equation}
where $\dataMat''[q,d] = \strange[q] \dataMat'[q,d] = \strange[q] \gamma \dataMat[q,d] \in \mathds{C}$ typically lies outside $\dataConst$.
Note that, in this case, the noise variance remains unchanged, and $\dataObs[n,q,d] \sim \CN(\staoa[n] \dataMat''[q,d], \noiseVar)$.

\subsubsection{Least Squares}
Based on these reformulations, the \gls{ml} problem becomes
\begingroup \small
\begin{equation}\label{eq:ML_sep_2}
         \UErangeEst, \UEaoaEst  = \argmax_{\UErangeTest, \UEaoaTest} \; \;
        \max_{\channelcoeffTest \in \mathds{C}} \; \mathcal{L}(\channelMatEst; \UErangeTest, \UEaoaTest, \channelcoeffTest) \\ 
        \max_{\dataTest'' \in  \mathds{C}^{\numQ \times \numD}}  \mathcal{L}(\dataObs ; \UEaoaTest, \dataTest''),
\end{equation}
\endgroup
where both the pilot and data parts are now affected by their own independent nuisance parameter.
Taking the logarithm of \eqref{eq:ML_sep_2} yields the following \gls{ls} formulation:
\begingroup \small
\begin{align}
    & \UErangeEst, \UEaoaEst =  \argmin_{\UErangeTest, \UEaoaTest} \; \min_{\channelcoeffTest \in \mathds{C}} \; 
        \numP \norm{
         \channelVecEst - \channelcoeffTest \stcomb(\UErangeTest, \UEaoaTest)}^{2}
          \nonumber \\
         & +   \min_{\dataTest'' \in  \mathds{C}^{\numQ \times \numD}} 
         \sum_{q=0}^{\numQ-1} \sum_{d=0}^{\numD-1}
         \norm{\dataObs[:,q,d] - \staoa(\UEaoaTest) \dataTest''[q,d]}^{2}, \label{eq:LS}
\end{align}
\endgroup
where the summations in the second term arise from the independence of the transmitted symbols and observations across the time and frequency dimensions.
From Wirtinger calculus \cite{koor_short_2023}, and given that the problem is separable in $q$ and $d$ with respect to $\dataTest''$, the solution to \eqref{eq:LS} for the nuisance parameters is
\begingroup \footnotesize
\begin{align}
& \channelcoeffEst(\UErangeTest, \UEaoaTest)  =  \frac{\stcomb^{H} (\UErangeTest, \UEaoaTest)}{\norm{\stcomb(\UErangeTest, \UEaoaTest)}^{2}} \channelVecEst = \frac{1}{\numRX\numQ} \stcomb^{H} (\UErangeTest, \UEaoaTest)\channelVecEst, \\
& \dataEst''[q,d](\UEaoaTest) = \frac{\staoa^{H}(\UEaoaTest)}{\norm{\staoa(\UEaoaTest)}^{2}} \dataObs[:,q,d] 
            = \frac{1}{\numRX} \staoa^{H}(\UEaoaTest) \dataObs[:,q,d].
\end{align}
\endgroup

By defining the projections matrices $\boldsymbol{P}_{\staoa}(\UEaoaTest) \triangleq \frac{\staoa(\UEaoaTest) \staoa^{H}(\UEaoaTest)}{\norm{\staoa(\UEaoaTest)}^{2}}$, $\boldsymbol{P}_{\stcomb}(\UErangeTest,\UEaoaTest) \triangleq \frac{\stcomb(\UErangeTest, \UEaoaTest) \stcomb^{H} (\UErangeTest, \UEaoaTest)}{\norm{\stcomb(\UErangeTest, \UEaoaTest)}^{2}}$, and exploiting the projection matrix properties ($\boldsymbol{P}^{H} = \boldsymbol{P}$ and $\boldsymbol{P}\boldsymbol{P}  = \boldsymbol{P}$),
we can show that 
\begingroup \footnotesize
\begin{align}
    & \norm{
    \channelVecEst - \channelcoeffEst(\UErangeTest, \UEaoaTest) \stcomb(\UErangeTest, \UEaoaTest)}^{2}  = \norm{\channelVecEst}^2 - \norm{
     \boldsymbol{P}_{\stcomb}(\UErangeTest, \UEaoaTest) \channelVecEst}^{2}, \label{eq:LS_pilot_Proj}\\
    & \norm{\dataObs[:,q,d] - \staoa(\UEaoaTest) \dataEst''[q,d](\UEaoaTest)}^2   = 
    \norm{\dataObs[:,q,d]}^2 \nonumber \\
    & \hspace{4cm} - \norm{\boldsymbol{P}_{\staoa}(\UEaoaTest) \dataObs[:,q,d]}^2. \label{eq:LS_data_Proj}
\end{align}
\endgroup

At this point, the dependency on the nuisance parameters has been effectively integrated through implicit estimations, leading to projection matrices. 
Applying the same properties, the second terms in \eqref{eq:LS_pilot_Proj} and \eqref{eq:LS_data_Proj} reduce, respectively, to:
\begingroup \small
\begin{align}
     & \norm{\boldsymbol{P}_{\stcomb}(\UErangeTest, \UEaoaTest) \channelVecEst}^{2} =  \channelVecEst^H \boldsymbol{P}^{H}_{\stcomb}(\UErangeTest, \UEaoaTest) \boldsymbol{P}_{\stcomb}(\UErangeTest, \UEaoaTest) \channelVecEst  \nonumber\\
    & = \channelVecEst^H  \boldsymbol{P}_{\stcomb}(\UErangeTest, \UEaoaTest) \channelVecEst = \frac{1}{\numRX \numQ} \abs{\channelVecEst^H  \stcomb(\UErangeTest, \UEaoaTest)}^2, \\
    & \norm{\boldsymbol{P}_{\staoa}(\UEaoaTest) \dataObs[:,q,d]}^2 = \dataObs[:,q,d]^H \boldsymbol{P}^{H}_{\staoa}(\UEaoaTest) \boldsymbol{P}_{\staoa}(\UEaoaTest) \dataObs[:,q,d] \nonumber \\
    & = \dataObs[:,q,d]^H  \boldsymbol{P}_{\staoa}(\UEaoaTest) \dataObs[:,q,d] = \frac{1}{\numRX} \abs{\dataObs[:,q,d]^H \staoa(\UEaoaTest)}^2.
\end{align}
\endgroup 
Thus, the final \gls{ls} formulation is given by
\begin{align}
    \UErangeEst, \UEaoaEst = \argmax_{\UErangeTest, \UEaoaTest} \; & 
    \frac{\numP}{\numRX \numQ} \abs{\channelVecEst^H  \stcomb(\UErangeTest, \UEaoaTest)}^2 \nonumber \\
    + &  \frac{1}{\numRX} \sum_{q=0}^{\numQ-1} \sum_{d=0}^{\numD-1} \abs{\dataObs[:,q,d]^H \staoa(\UEaoaTest)}^2 . \label{eq:PROJ_LS}
\end{align}

A few important remarks can be made regarding \eqref{eq:PROJ_LS}:
\begin{itemize}
    \item The pilot part provides both \gls{aoa} and range information through $\numRX\numQ$ coherent observations, while the data part provides only \gls{aoa} information through $\numRX$ coherent observations per symbol, with the $\numQ$ subcarriers contributing solely to \gls{snr}.
    \item Since $\channelcoeff$ is constant across the $\numP$ pilot symbols, these observations are combined in $\channelVecEst$, whereas the $\numD$ independent data observations contribute again only to \gls{snr}. 
    \item Through constellation relaxation, the resulting estimator is \textbf{constellation-agnostic}; regardless of the chosen constellations, the estimator and its performance remain unchanged.
\end{itemize}
\begin{figure*}[!b]
    \begin{align}
    \UErangeEst, \UEaoaEst = \argmax_{\UErangeTest, \UEaoaTest} \;  
    \frac{\numP}{\numRX\numQ}  \abs{
    \mathcal{F}_{\mathrm{2D}} \left\{ \channelMatEst^{H} \right\}
    \left( \frac{\numQ \Qspacing}{\carrierF} \frac{\UErangeTest}{\carrierWl}, \frac{\numRX \RXspacing}{\carrierWl} \sin(\UEaoaTest) \right)
    }^2 
    +  \frac{1}{\numRX} 
    \sum_{q=0}^{\numQ-1} \sum_{d=0}^{\numD-1}
     \abs{
    \mathcal{F}_{\mathrm{1D}} \left\{\dataObs^{*}[:,q,d]  \right\}
    \left(\frac{\numRX \RXspacing}{\carrierWl} \sin(\UEaoaTest) \right)
    }^2
     \label{eq:FFT}
\end{align}
\end{figure*}

\subsection{FFT-based Acceleration}
Solving \eqref{eq:PROJ_LS} requires evaluating the expression for all possible $(\UErangeTest, \UEaoaTest)$. 
By leveraging the linear phase shifts in $\stcomb(\UErangeTest, \UEaoaTest)$ and $\staoa(\UEaoaTest)$ arising from the \gls{ula} and \gls{ff} conditions, \eqref{eq:PROJ_LS} reduces to maximizing the sum of two \gls{fft} squared norms, which is expressed by \eqref{eq:FFT}.
The 1D and 2D Fourier Transform operators are denoted as $\mathcal{F}_{\mathrm{1D}} \left\{ \cdot \right\}$ and $\mathcal{F}_{\mathrm{2D}} \left\{ \cdot \right\}$, respectively. The resulting vector and matrix in \eqref{eq:FFT} are evaluated at samples corresponding to spatial frequencies $\frac{\numQ \Qspacing}{\carrierF} \frac{\UErangeTest}{\carrierWl}$ for range and $\frac{\numRX \RXspacing}{\carrierWl} \sin(\UEaoaTest)$ for \gls{aoa}.
This expression enables efficient computation when $\numRX$ and $\numQ$ are powers of two—which is the standard design—thanks to the \gls{fft} implementation.
Note that a detailed complexity analysis is provided in Section~\ref{sec:complexity}.

From \eqref{eq:FFT}, the expected classical properties in radar theory \cite{richards_principles_2010} can be observed:
\begin{itemize}
    \item The range resolution is given by $\UErangeRes = \frac{\carrierF}{\numQ\Qspacing}$ ($\carrierWl$) $= \frac{c}{\numQ\Qspacing}$ (\si{\meter}). 
    The maximum unambiguous range is \mbox{$R_{\mathrm{max}} = \frac{1}{2}\frac{\carrierF}{\Qspacing}$ ($\carrierWl$)}.
    \item The \gls{aoa} resolution in $\sin(\UEaoa)$ is given by $\UEsinaoaRes = \frac{\carrierWl}{\numRX\RXspacing}$, which simplifies to $\UEsinaoaRes = \frac{2}{\numRX}$ for the typical case $\RXspacing = \carrierWl/2$. In this case, all $\UEaoa \in \left]-\frac{\pi}{2}, \frac{\pi}{2}\right[$ can be estimated unambiguously.
\end{itemize}

Finally, after obtaining coarse estimates on the grid defined by the FFT, an optimization algorithm (here, Powell's method \cite{powell_efficient_1964}) is applied to the explicit form of the log-likelihood to refine the initial estimates.

\section{Numerical Results}
In this section, Monte Carlo simulations are performed to assess the localization accuracy improvement provided by the proposed method (denoted \textcolor{myGreen}{\underline{\projMethod}} here, referring to the paper title) against the following traditional baselines:

\textcolor{myRed}{\underline{\pilotMethod}}:
The simplest estimator uses only the pilot component and completely neglects the data part for localization.
This considers only the first term of \eqref{eq:FFT} and provides a lower bound on performance.

\textcolor{myBlue}{\underline{\pilotdataMethod}}:
The best theoretically achievable performance is obtained by assuming the data are perfectly known.
This estimator treats all $\numP+\numD$ symbols as pilots. 

\textcolor{myOrange}{\underline{\ddlmmseMethod}}:
The \gls{lmmse}-based \gls{dd} approach first estimates the channel and then demodulates the symbols, both according to the \gls{lmmse} criterion. 
The position is then retrieved using \pilotdataMethod, replacing the ground-truth data symbols with their estimated values—i.e., hard decisions.
Note that a \gls{zf}-based \gls{dd} has also been implemented but provides results very similar to, though slightly worse than, the \gls{lmmse} approach and is therefore not shown here for graph readability. 
Finally, note that this \gls{dd} method operates directly on symbols and does not account for error-correction coding, since we assume the opportunistic system either lacks access to the coding information or prefers to avoid performing such computationally intensive decoding. 

To evaluate the localization performance, we use the Hit-Rate and the \gls{rmse} for both $\sin(\UEaoaEst)$ and $\UErangeEst$. 
The HitRate is defined as the proportion of estimates falling within the main lobe of the likelihood, i.e., where the absolute error is smaller than half the resolution $\UEsinaoaRes$ or $\UErangeRes$.
Those two quantities are represented against the \gls{snr}, defined here as 
\begin{equation}
    \snr = \frac{\mathds{E}_{\channelcoeff}\{\abs{\channelcoeff}^2\} \dataConstVar}{\noiseVar},
\end{equation}
where $\dataConstVar$ is the symbol variance (assumed identical for pilots and data). 
From Assumptions \ref{ass:stillness} and \ref{ass:los}, no fading is modeled\footnote{For the sake of simplicity, we do not introduce either path-loss attenuation in the channel model or in the \gls{snr} definition, as it affects all methods identically. Nevertheless, it was verified through simulations that including it only shifts the \gls{snr} axis and does not change any conclusions regarding the results.} and $\channelcoeff = e^{j \phi}$ with $\phi \sim \mathcal{U}_{[0,2\pi)}$.
We set $\dataConstVar = 1$ and therefore, $\snr = 1/\noiseVar$.
Finally, each Monte Carlo trial draws random \gls{ue} position, data sequences, channel coefficients, and noise realizations.

\subsection{Performance Analysis}
\begin{figure*}[!t]
    \centering
    \begin{subfigure}[]{0.95\linewidth}
        \includegraphics[width=\linewidth]{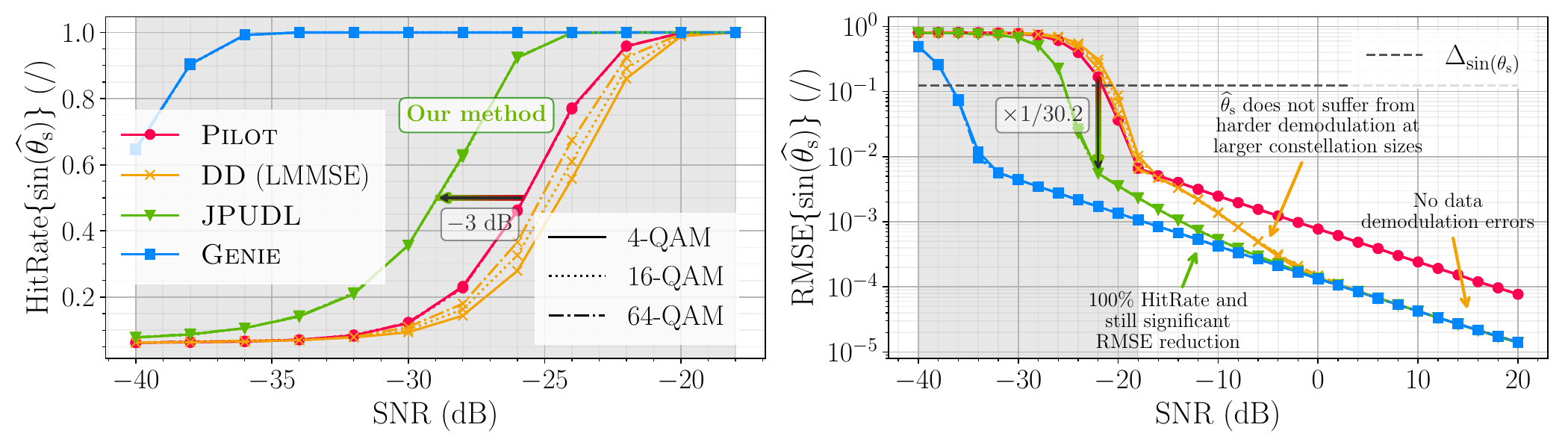}
        \caption{AoA results}
        \label{fig:results_sin_AoA}
    \end{subfigure}
    \\
    \begin{subfigure}[]{0.95\linewidth}
        \includegraphics[width=\linewidth]{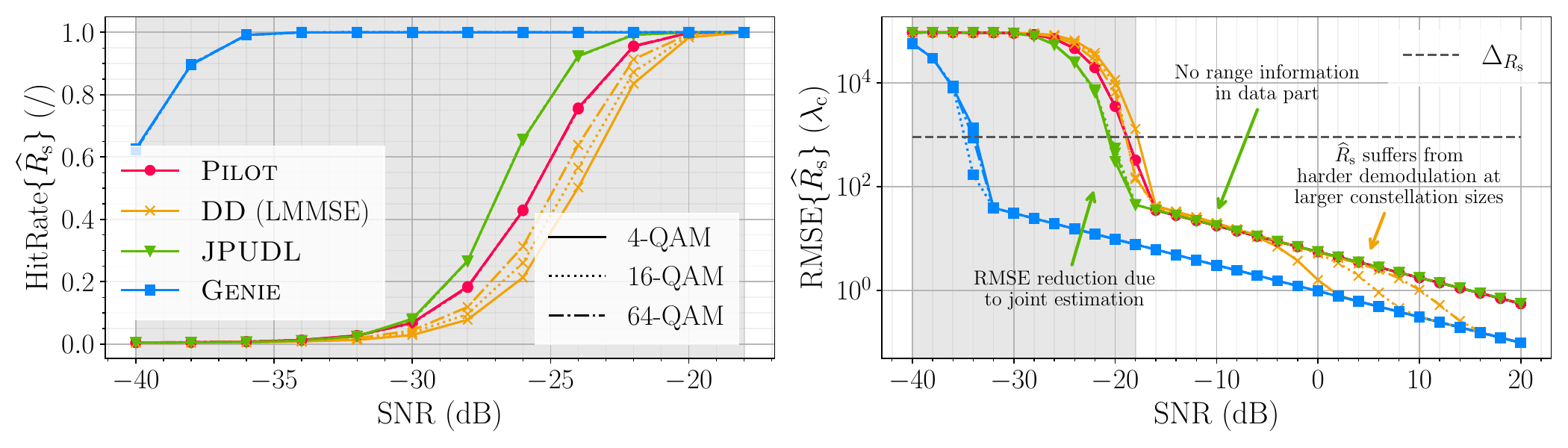}
        \caption{Range results (\gls{rmse} normalized by $\carrierWl$)}
        \label{fig:results_range}
    \end{subfigure}
    \caption{HitRate and RMSE of $\sin(\UEaoaEst)$ and $\UErangeEst$ as a function of SNR. 
    The HitRate is shown for $\snr \in [\SI{-40}{\decibel},\SI{-18}{\decibel}]$ (gray area) and the RMSE for $\snr \in [\SI{-40}{\decibel},\SI{20}{\decibel}]$.
    System parameters: $\numP=1$ (\texttt{BPSK} symbols), $\numD=32$, $\dataConst = \{4\texttt{QAM}\}$, $\{16\texttt{QAM}\}$ and $\{64\texttt{QAM}\}$, $\numQ=256$, $\Qspacing=\SI{15}{\kilo\hertz}$, $\numRX=16$, $\RXspacing=\SI{0.5}{\carrierWl}$. Results are obtained for 40000 Monte Carlo iterations.}
    \label{fig:results}
\end{figure*}

Figures \ref{fig:results} (\subref{fig:results_sin_AoA}) and (\subref{fig:results_range}) depict the results for the set of parameters given in the caption\footnote{Note that the HitRate is shown over a limited \gls{snr} range for readability, as all methods achieve \SI{100}{\percent} HitRate at higher \gls{snr} values.}.

As expected, our method significantly improves the \gls{aoa} estimation compared to pilot-based and \gls{dd} approaches.
This improvement is clearly visible on the HitRate curve, where a shift of up to \SI{3}{\decibel} in \gls{snr} is observed. 
It is also evident on the \gls{rmse} curve: a reduction in error by more than a factor of $30$ is achieved at $\snr = \SI{-22}{\decibel}$ when more than \SI{95}{\percent} of the estimates hit, and a reduction of around a factor of $4$ is obtained when all estimates hit at a \gls{snr} of \SI{-10}{\decibel}.
At higher \gls{snr} values, when all data symbols are correctly estimated, the \gls{dd} approach reaches optimal performance. 

Regarding range estimation, as expected at high \glspl{snr}, the data term does not provide information about $\UErange$ and our method performs similarly to \pilotMethod.
More surprisingly, at low \gls{snr} values (between $\SI{-30}{\decibel}$ and $\SI{-20}{\decibel}$), the range estimation is more accurate as it benefits from improved \gls{aoa} estimates during the joint optimization. 
A shift of \SI{1}{\decibel} in \gls{snr} is notably observed on the HitRate curve.

The reported gains hold for all constellation sizes considered.
As expected, the proposed method is \textbf{constellation-agnostic}.
In contrast, the \gls{dd} approach shows degraded range estimation for larger constellations (the \gls{rmse} drop shifts to higher \gls{snr}), while \gls{aoa} estimation remains unaffected.
This asymmetry arises because \gls{aoa} is derived from phase differences across antennas, which are independent of the transmitted symbols, whereas range relies on phase differences across subcarriers, where different communication symbols are sent, hence requiring correct demodulation to guide the estimator toward the true value.

Another observation can be drawn: at \gls{snr} values below \SI{-20}{\decibel}, the \gls{dd} approach performs worse than the pilot-based method. 
This can be explained by the fact that data detection errors become so significant that they degrade the equalized observations $\channelVecEst$ when using the full sequence of $\numP + \numD$ symbols compared to using only the $\numP$ pilots.
Additionally, at very low \glspl{snr}, the \gls{dd} approach counterintuitively performs better with larger constellations; this can be attributed to the denser symbol constellation offering a higher probability that an incorrect symbol lies closer to the true one.

Finally, our method consistently outperforms pilot-based and \gls{dd} approaches for $\UEaoaEst$, while improving—at low \gls{snr} values—or at least preserving $\UErangeEst$ performance, except when data detection is nearly perfect.
However, this condition is not guaranteed and is often rare in opportunistic radar configurations.
Moreover, the \gls{pr} is not inherently concerned with data detection, and including it introduces additional computational overhead, as discussed below.
In contrast, our method eliminates the need for data detection and remains computationally efficient through the \gls{fft} implementation—an important advantage for opportunistic systems, where minimizing energy consumption (and hence computational requirements) is critical.
A detailed comparison of the computational requirements of the proposed method against the \gls{dd} approach is provided in Section~\ref{sec:complexity}.

\subsection{Influence of System Parameters}

This section analyzes the effect of system parameters on localization performance.

\begin{figure*}[!ht]
    \centering
    \begin{subfigure}[]{0.45\linewidth}
        \centering
        \includegraphics[width=\linewidth]{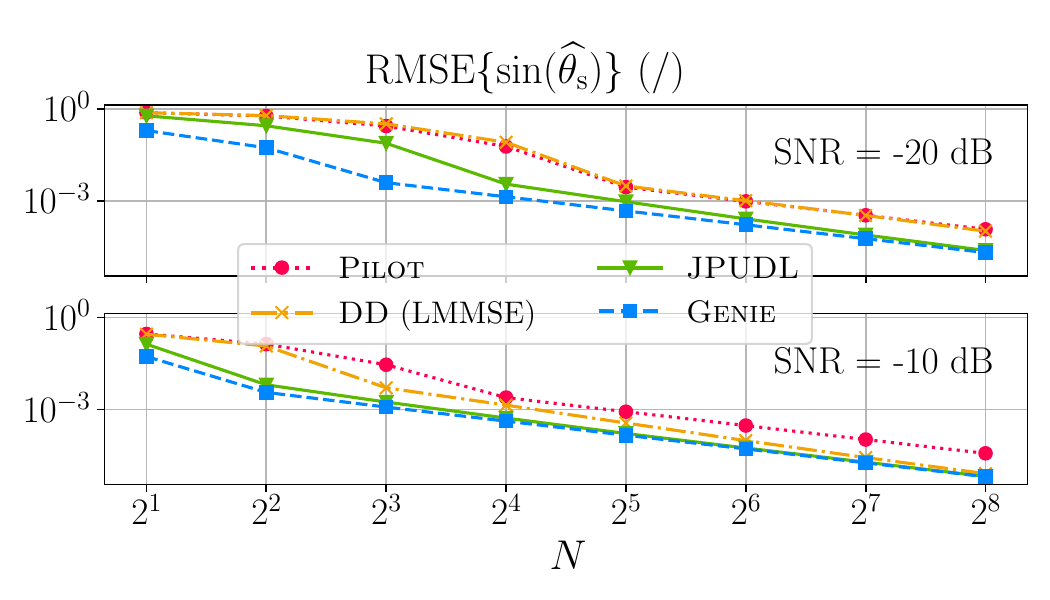}
        \caption{\gls{rmse} of $\sin(\UEaoaEst)$ as a function of $\numRX$ and for two \glspl{snr}.}
        \label{fig:results_sweep_N}
    \end{subfigure}
    \begin{subfigure}[]{0.45\linewidth}
        \centering
        \includegraphics[width=\linewidth]{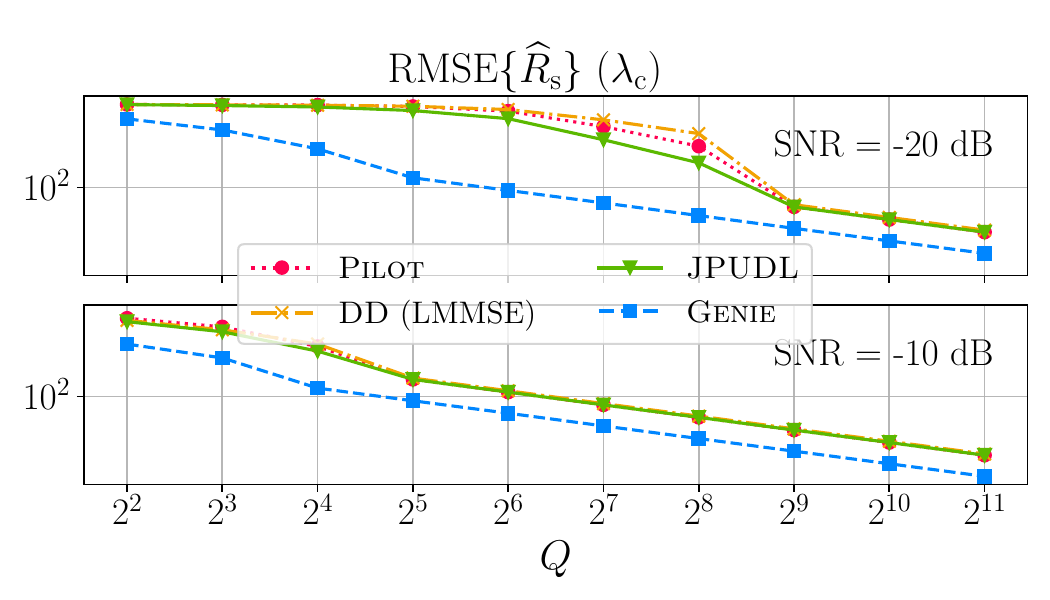}
        \caption{\gls{rmse} of $\UErangeEst$ as a function of $\numQ$ and for two \glspl{snr}.}
        \label{fig:results_sweep_Q}
    \end{subfigure}
    \caption{System parameters: $\numP=1$ (\texttt{BPSK} symbols), $\numD=32$, $\dataConst = \{16\texttt{QAM}\}$, $\numQ=256$ for (a), $\Qspacing=\SI{15}{\kilo\hertz}$, $\RXspacing=\SI{0.5}{\carrierWl}$, $\numRX=16$ for (b). Results are obtained for 10000 Monte Carlo iterations.}
    \label{fig:results_sweep}
\end{figure*}

\autoref{fig:results_sweep} (\subref{fig:results_sweep_N}) depicts the \gls{rmse} of $\sin(\UEaoaEst)$ as a function of $\numRX$ for \gls{snr} values of \SI{-20}{\decibel} and \SI{-10}{\decibel}.
As expected, increasing the number of \gls{rx} antennas improves \gls{aoa} estimation, owing to the augmented steering vector $\staoa$ and the finer resolution $\UEsinaoaRes$.
For instance, the \gls{rmse} decreases from $1$ to $10^{-3}$ as $\numRX$ increases from $2$ to $32$ at \SI{-20}{\decibel} for \textcolor{myGreen}{\projMethod}.
Although not shown here for brevity, note that improved \gls{aoa} also benefits range estimation, since joint optimization is performed.

\autoref{fig:results_sweep} (\subref{fig:results_sweep_Q}) illustrates the \gls{rmse} of $\UErangeEst$ as a function of $\numQ$ for \gls{snr} values of \SI{-20}{\decibel} and \SI{-10}{\decibel}.
As anticipated, increasing the number of subcarriers $\numQ$ improves range estimation.
Again, this also enhances $\UEaoaEst$.
Range estimation benefits from the enriched steering vector $\strange$, but also from the intrinsic range resolution $\UErangeRes$.
This resolution can be further improved (i.e., $\UErangeRes$ reduced) without requiring additional observations along $q$ by increasing the subcarrier spacing $\Qspacing$.
This translates into better range accuracy, provided the coarse grid resolution is sufficiently fine relative to $\UErangeRes$ to capture the main lobe of the objective function.

Finally, note that, quite naturally, increasing $\numP$ or $\numD$ also improves the overall localization performance.

\newpage
\section{Complexity Analysis}\label{sec:complexity}

This section analyzes the complexity of the proposed method against the considered baselines.
\autoref{tab:complexity} reports the asymptotic computational complexity of the processing steps for each method.
The first column corresponds to the \gls{ls} formulation \eqref{eq:PROJ_LS}, evaluated over a 2D grid of $\numRXgrid\numQgrid$ points in $(\UEaoaTest, \UErangeTest)$ for the pilot term and a 1D grid of $\numRXgrid$ points in $\UEaoaTest$ for the data term.
The second column represents the \gls{fft}-based implementation of the proposed method \eqref{eq:FFT}, with \gls{fft} sizes matched to the respective grid sizes\footnote{
After grid evaluation, estimates are refined via an optimization algorithm requiring additional objective function evaluations. 
Since the amount of such evaluations depends on the solver strategy rather than on the proposed estimators, it is excluded from the complexity analysis. 
In practice, a sufficiently fine search grid yields adequate localization accuracy without refinement.
} for a fair comparison, yielding reductions of $\log(\numRXgrid\numQgrid) \ll \numRX\numQ$ and $\log(\numRXgrid) \ll \numRX$.
The last three columns correspond to the considered baselines, all using the efficient \gls{fft}-based implementation.
Compared to the baselines, the proposed method incurs an additional cost of $\Compl{\numRXgrid \log(\numRXgrid)\numQ \numD}$ for the $\numQ \numD$ 1D \glspl{fft} on $\channelMatEst$.
On the other hand, the \gls{dd} baseline introduces an overhead of $\Compl{\numRX \numQ \numD + M \numQ \numD }$ for computing hard data decisions.
As expected, the pilot-only approach is always faster than the proposed method, as exploiting data payloads for positioning comes at an unavoidable computational cost. 
However, the proposed method is faster than \gls{dd} for large constellations or restricted grid sizes, trading $\Compl{\numRX\numQ \numD + M \numQ \numD}$ for $\Compl{\numRXgrid \log(\numRXgrid) \numQ  \numD}$.

\begin{table*}
    \centering
    \caption{Asymptotic Computational Complexity per Processing Step}
    \label{tab:complexity}
    \begin{tabular}{|l||c|c|c|c|c|}
        \hline
        Step & \projMethod \,\eqref{eq:PROJ_LS} & \textcolor{myGreen}{\projMethod} \eqref{eq:FFT} & \textcolor{myRed}{\pilotMethod} & \textcolor{myBlue}{\pilotdataMethod} & \textcolor{myOrange}{\ddlmmseMethod}   \\
        \hline
        Channel estimation  & — & — & — & — & $\Compl{\numRX \numQ \numP}$ \\
        \hline 
        Soft data estimation & — & — & — & — & $\Compl{\numRX \numQ \numD }$ \\
        \hline
        \makecell[l]{Hard data decision \\ ($M$-ary constellation)}  & — & — & — & — & $\Compl{M \numQ \numD}$ \\
        \hline 
        \makecell[l]{Construction of $\channelVecEst$} & $\Compl{\numRX \numQ \numP}$ & $\Compl{\numRX \numQ \numP}$ & $\Compl{\numRX \numQ \numP}$ & $\Compl{\numRX  \numQ (\numP+\numD)}$ & $\Compl{\numRX \numQ (\numP+\numD)}$  \\
        \hline 
        Function evaluation &  \makecell[c]{$\Compl{\numRXgrid\numQgrid \numRX \numQ}$ \\ $+ \Compl{\numRXgrid \numRX \numQ \numD}$} &  \makecell[c]{$\Compl{\numRXgrid\numQgrid \log (\numRXgrid\numQgrid)}$ \\ $ +  \Compl{\numRXgrid \log(\numRXgrid)  \numQ \numD}$} & $\Compl{\numRXgrid\numQgrid \log (\numRXgrid\numQgrid)}$ & $\Compl{\numRXgrid\numQgrid \log (\numRXgrid\numQgrid)}$ & $\Compl{\numRXgrid\numQgrid \log (\numRXgrid\numQgrid)}$ \\
        \hline
    \end{tabular}
\end{table*}

\section{Conclusion}

This paper investigates the joint exploitation of pilots and unknown data payloads for localization. 
We propose a novel method that does not require data detection, making it robust under poor communication conditions (i.e., at low \gls{snr}). 
The method is derived for an uplink scenario with a \gls{ula} at the \gls{srx}.
In this configuration, the estimator is shown to be computationally efficient via an \gls{fft}-based implementation.
Through numerical simulations, we demonstrate that our method achieves superior localization performance compared to existing approaches in an opportunistic scenario, where no communication performance guarantees are available.
Additionally, we analyze the impact of system parameters on both localization performance and computational requirements.
Ongoing work extends this methodology to \gls{nf} localization with distributed antenna arrays.
Future work may consider multipath propagation, alternative configurations such as multistatic radar relying on signal echoes, and extensions to multi-user scenarios, as well as imperfect synchronization.


\newpage
\bibliographystyle{IEEEtran}
\bibliography{nourl,references}

\end{document}